\documentclass[12pt]{article}

\usepackage[body={17.5cm, 22cm},right=2cm]{geometry}
\usepackage{amssymb}
\usepackage{amsmath}

\def\G{{\mathcal G}}

\def\l{\left}
\def\r{\right}

\def\beq{\begin{equation}}
\def\eeq{\end{equation}}
\def\bea{\begin{eqnarray}}
\def\eea{\end{eqnarray}}
\def\barr{\begin{array}}
\def\earr{\end{array}}
\def\be{\begin{equation}}
\def\ee{\end{equation}}
\def\bc{\begin{center}}
\def\ec{\end{center}}

\def\ov{\overline}

\def\Ne#1{{\cal N}=#1}
\def\Ng#1{{\cal N}>#1}

\def\de{\partial}

\def\Im{\mathop{\mbox{Im}}}
\def\Re{\mathop{\mbox{Re}}}

\begin{document}
\setcounter{page}{0}
\thispagestyle{empty}

\begin{titlepage}
\begin{flushright}
DFPD-2011/TH/15 \\ 
SLAC-PUB-14627
\end{flushright}

~\vspace{1cm}
\begin{center}

{\LARGE \bf On Fayet-Iliopoulos terms and de~Sitter \\[.4cm] vacua in supergravity: some easy pieces}

\vspace{1.2cm}

{\large \bf
Francesca Catino$^{a}$,
Giovanni Villadoro$^{b}$, 
Fabio Zwirner$^{a}$}
\\
\vspace{.6cm}
{\normalsize {\sl $^a$ Dipartimento di Fisica, Universit\`a di Padova and INFN, Sezione di Padova, 
\\ Via Marzolo 8, I-35131 Padova, Italy}}

\vspace{.3cm}
{\normalsize { \sl $^{b}$ SLAC, Stanford University \\
2575 Sand Hill Rd., Menlo Park, CA 94025 USA}}

\end{center}
\vspace{.8cm}
\begin{abstract}
We clarify a number of issues on Fayet-Iliopoulos (FI) terms in supergravity, keeping the formalism at a minumum and making use of explicit examples. We explain why, if the $U(1)$ vector is massive everywhere in field space, FI terms are not genuine and can always be redefined away or introduced when they are not present. We formulate a simple anomaly-free model with a genuine FI term, a classically stable de Sitter (dS) vacuum and no global symmetries. We explore the relation between $\Ne2$ and $\Ne1$ FI terms by discussing $\Ne1$ truncations of $\Ne2$ models with classically stable dS vacua.
\end{abstract}

\end{titlepage}

%


\tableofcontents
\vspace{1cm}
\section{Introduction and conclusions}
\label{intro}

After several decades of extensive studies on simple and extended four-dimensional supergravities, and on the r\^ole of these theories as effective low-energy theories of superstring or M-theory compactifications, the theoretical status of Fayet-Iliopoulos (FI) terms \cite{FIglobal} in supergravity \cite{FIlocal} and its possible ultraviolet completions is still under discussion (for some recent literature, see e.g. \cite{recent, dt, FIquant}).

Several $\Ne1$ supergravity models with a gauged $U(1)$ R-symmetry, associated with a constant FI term, have been formulated \cite{models}, typically leading to Minkowski or de Sitter (dS) vacua with a massive vector boson associated to the spontaneously broken $U(1)$ R-symmetry. Quantization of the FI parameter in supergravity was inferred in \cite{Witten:1985bz} and discussed in more detail in \cite{FIquant}. 

In this paper we clarify some pending issues, keeping the formalism at a minimum and using a number of explicit models for illustration.

In section~2 we first recall some well known facts about FI terms in $\Ne1$ supergravity, also to introduce our notation. We do not consider the so-called field-dependent FI terms, which do not involve the gauging of an R-symmetry and are nothing else than $U(1)$ D-terms for a gauge symmetry that does not act linearly on the fields. We concentrate instead on constant FI terms, associated with the gauging of an R-symmetry, and explain how we should distinguish between genuine FI terms and impostors. Whenever there is no point in field space where the $U(1)$ R-symmetry is restored, the FI term can be shifted by an arbitrary amount, in particular it can be defined away if present or introduced if absent, without changing the physical content of the theory. Such FI terms are thus impostors, and cannot obey a quantization condition. This point is a direct consequence of the general structure of supergravity and may be known to some\footnote{See, e.g., the example discussed in section~4.5 of \cite{dt}, which is closely related to the example we will discuss in section~3.1. We thank K.~Dienes and B.~Thomas for bringing their example to our attention.}, but to the best of our knowledge an explicit and general supergravity formulation was never given in the literature. We conclude the section by recalling the anomaly cancellation conditions, which are essential for quantum consistency.

In section~3 we present two simple examples. First, we display a model with a FI impostor. We start from the theory of a free massive $U(1)$ vector superfield, with no FI term, and write down explicitly the analytic chiral superfield redefinition and K\"ahler transformation that map the theory into an equivalent one with an impostor FI term. Second, we present an anomaly-free model with a genuine FI term, a classically stable dS vacuum and no exact global symmetry. Depending on the choice of the parameters, the vector field can be either massless or massive on the vacuum. The scalar potential is positive definite and supersymmetry is always broken, except when the superpotential is trivial. In that case the vacuum energy vanishes and the FI term induces a supersymmetric mass for the vector supermultiplet. In the model, all physical masses and energy densities can be made parametrically small with respect to the Planck scale, even if the FI term is assumed to be quantized in Planck mass units. Interestingly, the relevant mass scale of the model saturates the bound from the weak gravity conjecture of ref.~\cite{ArkaniHamed:2006dz}.

Objects analogous to the $\Ne1$ FI terms do also exist in $\Ne2$ \cite{N2FI} and $\Ne4$ \cite{N4FI} supergravity: they play a crucial r\^ole in the construction of the only known classically stable dS vacua in extended supergravity \cite{N2FI, altri}, whilst no stable dS vacuum has been found so far in $\Ng2$ supergravity. In section 4, we discuss two consistent $\Ne1$ truncations of the simplest $\Ne2$ model of \cite{N2FI}, which contains three vector multiplets and no hypermultiplet, and gauges $SO(2,1) \times U(1)$, with the $\Ne2$ FI term associated with the $U(1)$ factor\footnote{Complementary considerations on $\Ne2$ FI terms and $\Ne1$ truncations can be found in \cite{achuc}.}. In one case, the $U(1)$ part of the $\Ne2$ potential is mapped into the $\Ne1$ F-term potential, whilst the truncation of the $SO(2,1)$ part of the $\Ne2$ potential generates a genuine FI-term contribution to the $\Ne1$ potential. In the other case, both contributions to the $\Ne2$ potential are mapped into $\Ne1$ potentials generated by FI terms. In both cases, the truncated $\Ne1$ theory is anomalous, but additional charged chiral multiplets can be added while keeping the same vacuum, in analogy with the twisted-sector fields of orbifold string constructions. In conclusion, the relation between the $\Ne2$ and the $\Ne1$ FI terms is not one-to-one but depends on the considered truncation. The general rule, valid also for other examples, is that the FI term of the truncated $\Ne1$ theory is associated, in the $\Ne2$ theory, to the linear combination of the generator of the compact subgroup of $SO(2,1)$, if gauged, and the component of the $\Ne2$ FI terms that survive the truncation, if any. On the other hand, the other components of the $\Ne2$ FI terms, if present, produce superpotential terms in the truncated theory.

\section{Known and less known facts on FI terms in supergravity}

\subsection{U(1) D~terms in $\Ne1$ supergravity}

The gauge-invariant two-derivative action for $\Ne1$, $D=4$
supergravity with chiral multiplets $\Phi^i \sim (z^i, \psi^i, F^i)$ and
vector multiplets $V^a \sim (\lambda^a, A_\mu^a, D^a)$ is completely 
fixed by three ingredients \cite{Dbooks}. The first is the real and
gauge-invariant K\"ahler function $G$, which can be written in terms
of a real K\"ahler potential $K$ and a holomorphic superpotential $W$
as\footnote{Here and in the following, we work in the natural units of supergravity, where the reduced Planck mass $M_P=1/\sqrt{8 \, \pi \, G_N}$ is set equal to one.}
\begin{equation} 
G = K + \log |W|^2 \, .  
\label{ggen}
\end{equation}
The second is the holomorphic gauge kinetic function $f_{ab}$, which transforms as a symmetric product of adjoint representations, plus a possible imaginary shift associated with anomaly cancellation. Generalized Chern-Simons terms may also be needed \cite{afl}, but they will not be relevant here, also because we will mostly focus on the simple case of a single Abelian gauge group factor. The third are the holomorphic Killing vectors $X_a = X_a^i (z) (\de/ \de z^i)$, which generate the analytic isometries of the K\"ahler manifold for the scalar fields that are gauged by the vector fields. In most of what follows it will suffice to think of $G$, $f_{ab}$ and $X_a$ as functions of  the complex scalars $z^i$ rather than the superfields $\Phi^i$ (as done, for example, in Appendix~G of \cite{Dbooks}). However, whenever needed we will turn to superfield notation.

The gauge transformation laws and covariant derivatives for the
scalars in the chiral multiplets read
\be
\delta z^i  =  X_a^i \, \epsilon^a \, , 
\qquad
D_\mu z^i = \de_\mu z^i - A^a_\mu X_a^i \, ,
\ee
where $\epsilon^a$ are real parameters. The scalar potential is
\be 
\label{vgen}
V = V_F + V_D = e^G \l (G^i G_i - 3 \r) + \frac12 D_a D^a \, ,
\ee
where $G_i = \de G / \de z^i$, scalar field indices are raised with
the inverse K\"ahler metric $G^{i \ov{k}}$, gauge indices are raised
with $[(Re f)^{-1}]^{ab}$, and 
\be
\label{eq:solD} 
D_a = i \, G_i \, X_a^i = i \, K_i \, X_a^i + 
i \, \frac{W_i}{W} \, X_a^i \, .  
\ee

Gauge invariance of $G$ requires that $K$ and $W$ be invariant up to a
K\"ahler transformation
\be
K' = K + H + \ov H \, ,
\qquad
W'  =  W \, e^{-H} \, ,
\ee
where $H$ is a holomorphic function, thus it will not be restrictive
to assume that $K$ is gauge invariant. If $W$ is also gauge-invariant,
eq.~(\ref{eq:solD}) reduces to the standard form
\be 
D_a = i \, K_i \, X_a^i \, .
\ee
For a linearly realized gauge symmetry, $$i \, K_i \, X_a^i = 
- K_i \, (T_a)^i_{\; k} z^k\,,$$ and we recover the standard
expression of \cite{cremmer} for the D-terms. 
For example, in the case of canonical K\"ahler potential,
and fields $z^i$ with definite charges $q^i$ with respect to a single U(1) gauge factor 
\begin{equation} \label{KXlinear}
K=\sum_{i}|z^i|^2\,,\qquad X^i=i \, q^i \, z^i\, ,
\end{equation}
and the $D$-term (with an implicit lower index) reads
\be
D=-\sum_i q^i\, |z^i|^2\,.
\ee
For an axionic realization, $X_a^i = i \, q_a^i$, where $q_a^i$ is a real constant,
and we obtain what are often called, with an abuse of language, field-dependent FI terms. 
A classic example \cite{dsw}, which often arises in string compactifications, is 
\begin{equation}\label{KXnlinear}
K=-\log(S+\overline S)\,, \qquad X^S = i \, q^S\,,
\end{equation}
which leads to
\be
D= \frac{q^S}{S+\overline S}\,.
\ee

If $W$ is not gauge invariant, it must be
\be \label{Xisuperpot}
i \, \frac{W_i}{W} \, X_a^i = \xi_a \, ,
\qquad
(\xi_a \in \mathbb{R}) \, ,
\ee
so that the gauge non-invariance of $W$ can be at most an overall
phase with real parameter $\xi_a$, for the Abelian factors
$U(1)_a$. The constants $\xi_a$ correspond to gaugings of
R-symmetries, and give rise to the supergravity expression for the
D-terms \cite{FIlocal, Dbooks}:
\be 
D_a = i \, K_i \, X_a^i + \xi_a \, .
\label{Dgen}
\ee
The $\xi_a$ are then the constant FI terms of supergravity, on which we will focus our attention from now on.
For example, in the linear case of eq.~(\ref{KXlinear}) and assuming $q^1=-\xi$, the superpotential
\begin{equation}
W=M^2\, z^1
\end{equation}
gauges a suitable U(1) R-symmetry, modifying the D-term into
\be
D=\xi-\sum_{i} q^i\, |z^i|^2\,.
\ee
Similarly,  in the non-linear case of eq.~(\ref{KXnlinear}) and assuming $q^S=\xi$, the superpotential  
\begin{equation}
W= W_0 \, e^{- S}  \, ,   
\end{equation}
where $W_0$ is a non-vanishing $S$-independent factor, also gauges a U(1) R-symmetry, and modifies the D-term into
\be
D= \xi \left( 1 + \frac{1}{S+\overline S} \right) \,.
\ee

To conclude, we recall the generic superfield expression of the $\Ne1$ supergravity Lagrangian with a gauged $U(1)$ R-symmetry\footnote{We neglect here additional gauge group factors, possible generalized Chern-Simons terms and constant numerical factors that are not important for the present considerations.}, in the compensator formalism (see, e.g. \cite{Kugo}):
\be
\label{complag}
{\cal L} = \left[ \, \overline{\Phi_0} \Phi_0 \, e^{-K/3} \right]_D 
+ \left( \left[ \Phi_0^3  \, W  \right]_F +  \left[ f  \, {\cal W} \, {\cal W} \right]_F + {\rm h.c.} \right) \, .
\ee
In the above expression: $\Phi_0$ is the chiral compensator superfield, transforming as $\Phi_0 \rightarrow \Phi_0^\prime = \Phi_0 \, \exp ( \xi \Lambda/3)$; $K = K_0 - \xi \, V$, where $K_0$  is a real and gauge-invariant function of the chiral superfields $\Phi$, of their conjugates $\overline{\Phi}$ and of the real gauge vector superfield $V$, the latter transforming as $V \rightarrow V' = V - \Lambda - \overline{\Lambda}$; $W$ is now an analytic function of the chiral superfields $\Phi$, transforming as $W \rightarrow W' = W \, \exp ( - \xi \Lambda)$;
$f$ is the gauge kinetic function (with implicit lower indices), analytic in the chiral superfields $\Phi$; ${\cal W}$ is the chiral supersymmetric field strength of $V$. Notice that having a FI term $\xi$ corresponds to giving charge $\xi/3$ to the compensator field $\Phi_0$ under the gauged $U(1)$ R-symmetry.

\subsection{Genuine FI terms and impostors}

In this section we show how, in theories where the gauge boson of the U(1) R-symmetry 
is massive everywhere in field space, the FI term associated with such vector field 
is not well defined and can always be redefined away---a genuine FI term, whose value 
cannot be shifted continuously, exists only when the theory allows the gauged $U(1)$ 
R-symmetry to be restored at least in one point in field space.

The general argument is quite simple and follows from the supergravity formalism reviewed  above.
The supermultiplet of a massive vector contains the degrees of freedom of a chiral supermultiplet besides those of a massless vector multiplet. In particular, the superfield of a massive vector $V_m$ can always be decomposed into
\begin{equation} \label{VmVS}
V_m=V+S+\overline{S}\,,
\end{equation}
with a massless vector superfield $V$ and a chiral superfield $S$ transforming as
\begin{equation} 
\label{SVtrans}
V \to V'=V-\Lambda-\overline{\Lambda} \, ,
\qquad
S\to S'=S+\Lambda \, .
\end{equation}
The l.h.s. of eq.~(\ref{VmVS}) can be thought of as the vector superfield in the unitary gauge $S=0$, while the r.h.s. is the gauge-invariant combination obtained from the unitary gauge via the St\"uckelberg trick.

In theories where the vector field is massive everywhere in field space, the field $S$ is globally defined,
since it corresponds to the superfield multiplet of the longitudinal mode. On the other hand, in theories where the gauge symmetry is restored somewhere in field space, $S$ is not globally defined. In the points where the vector mass vanishes, the would-be $S$ field is not dynamical (there is no kinetic term) 
and corresponds to a pure gauge.

Consider a supergravity model with a non-vanishing FI term $\xi$, associated to a vector superfield $V$ gauging an $R$-symmetry that is broken everywhere in field space. In this case there exists a globally defined chiral superfield $S$ transforming as in eq.~(\ref{SVtrans}). We can now perform a K\"ahler transformation using such field, namely
\begin{equation} \label{aStrans}
K'=K+\alpha \, (S+\overline S)\,, \qquad W'=We^{-\alpha \, S}
\end{equation}
with $\alpha$ an arbitrary real constant\footnote{In general, the K\"ahler transformation may be anomalous. However, we assume that the vector superfield $V$ is associated to an anomaly-free $U(1)$ R-symmetry. Such symmetry can be exploited to redefine the fields and make the full ``Kahler transformation + field redefinition'' anomaly-free.}. In terms of $K'$, the formula for the D~term (\ref{Dgen}) now reads
\begin{equation}
D=i K_i X^i+\xi = i K'_i X^i+\alpha+\xi\equiv i K'_i X^i+\xi'\,,
\end{equation}
from where we can see that in the new frame the FI term has been redefined. If we choose $\alpha=-\xi$, the FI term can be shifted away---in this theory the FI term is not well defined. We obtain the same conclusion by looking at the superpotential. Indeed the charge of the new superpotential $W'$ will be shifted into $-\xi-\alpha$, because of the extra contribution from the $S$ field. When $\alpha=-\xi$, $W'$ will be invariant, which again corresponds to no FI term\footnote{In the compensator formalism of eq.~(\ref{complag}), the K\"ahler fransformation (\ref{aStrans}) corresponds to the redefinition $\Phi_0\to\Phi_0'=\Phi_0 \, e^{-\alpha S/3}$, which shifts the compensator charge by $\alpha/3$ and the value of the FI term accordingly.}.

Notice that, since this FI term can be shifted by an arbitrary amount, it will not be subjected to quantization conditions. Since theories of this type are equivalent to one with a massive vector not associated to the R-symmetry, in this case we are not allowed to speak of genuine FI terms---such FI terms are impostors.

On the contrary, in theories where there are points in field space where the gauged R-symmetry is restored, we cannot use the field $S$ to redefine the FI term. Only in this case a proper FI term can be defined, we call such terms genuine FI terms.

\subsection{Anomaly cancellation conditions}

For a consistent effective theory, all gauge and gravitational
anomalies associated with a gauged $U(1)$ must vanish: in
particular, the cubic (${\cal A}_{U(1)^3}$), the gravitational (${\cal
A}_{U(1)}$) and the mixed-gauge anomaly (${\cal A}_{U(1)\,\G^2}$) if
the full gauge group is $U(1) \times \G$. 

To fix the notation, we assign the $U(1)$ charges as\footnote{ The standard
$R$-charge of the literature on global supersymmetry is $R = - 2 \, Q$.}
\be
\label{R-charges}
Q[\theta] =
Q[\lambda^a] = 
Q[\psi_\mu] = - \xi/2 \, , 
\quad
Q[W] = - \xi \, , 
\quad
Q[z^i] = q^i \, , 
\quad
Q[\psi^i] = q^i + \xi/2 \, , 
\ee
where $\theta$ is the anticommuting coordinate and $\psi_\mu$ is the gravitino.

In the above conventions, the fermionic contributions to the cubic and gravitational anomalies are:
\bea
Tr\,Q^3 & = & 3 \ (Q[\psi_\mu])^3
+ \sum_a (Q[\lambda^a])^3 + \sum_i (Q[\psi^i])^3 \, ,
\label{cubic}
\\ 
Tr \, Q & = & - 21 \  Q[\psi_\mu]
+ \sum_a Q[\lambda^a] + \sum_i  Q[\psi^i] \, ,
\label{linear}
\eea
see \cite{anom32} for the gravitino contributions. These contributions 
must either vanish or cancel possible Green-Schwarz (GS) contributions
\cite{gs} coming from the variation of $\Im f_{ab}$. All the resulting 
conditions are model dependent, in particular: all of them depend on 
the matter content; the GS contribution to ${\cal A}_{U(1)}$ depends on 
higher derivative terms ($R^2$); ${\cal A}_{U(1)\, \G^2}$ depends also on 
the details of $\G$ and its representations. However, there are in principle 
strong combined constraints on the possible matter content and on the 
$U(1)$ charges.

\section{Examples}

\subsection{A model with a FI impostor} 

Consider a free massive vector superfield  with the usual kinetic F-term and a mass term appearing in the K\"ahler potential as
\begin{equation}
K=\frac{1}{2} \, M^2 \, V_m^2 \qquad {\rm or} \qquad K = \frac{1}{2} \, (S+\overline S + MV)^2 \, ,
\end{equation}
where the first expression refers to the unitary gauge (a discussion of this model, in global supersymmetry and in the unitary gauge, can be found in section~4.5 of \cite{dt}), whilst the second uses the Wess-Zumino gauge for $V$, with the longitudinal components of the massive vector contained in the chiral superfield $S$. With respect to eq.~(\ref{VmVS}), we have reabsorbed the factor $M$ in $S$, to have the latter superfield canonically normalized. For definiteness, we can take a constant gauge kinetic function $f_0=1/g^2$ and a constant superpotential $W_0$, but we are allowed to take the limit $W_0 \rightarrow 0$ at the end of the calculations. Under gauge transformations:
\be
V \to V- \Lambda - \overline\Lambda \, ,
\qquad
S \to S+  M \, \Lambda \, .
\ee
The action described above corresponds to a massive Abelian vector superfield, with no FI term.

We now perform a trivial field redefinition of the chiral superfield $S$,
\begin{equation} 
\label{StoT}
S=T-\frac{\xi}{2 \, M} \, ,
\end{equation}
where $\xi$ is a real constant and $T$ transforms as $S$ under gauge transformations. The K\"ahler potential now reads
\begin{equation}
K= \frac{1}{2} \, (T+\overline{T}+M V)^2 -\frac{\xi}{M}(T+\overline{T}+MV)+\frac{\xi^2}{2 \, M^2} \, , 
\end{equation}
and after a K\"ahler transformation we have
\be
K= \frac{1}{2} \, (T+\overline{T}+M V)^2 -\xi V \, ,
\qquad
W= W_0 \, e^{\frac{\xi^2}{4 \, M^2}} e^{-\frac{\xi}{M} \, T} \, .
\ee
This is a theory of a massive vector superfield with a FI term, which consistently appears both as a linear term in $V$ in the K\"ahler potential and as a gauge non-invariance of the superpotential [see eq.~(\ref{Xisuperpot})]. As a check of the equivalence of the two theories, we can look at the expressions of the D~terms in the two frames, according to eq.~(\ref{eq:solD}):
\begin{align}
{\rm (i)} \qquad & D=i \, K_S \, X^S  = -  M \, (S+\overline S) \, ,  
\\
{\rm (ii)} \qquad & D=i \, K_T \, X^T +i \, \frac{W_T}{W} \, X^T  =\xi - M \, (T+\overline T) \, ,
\end{align}
which coincide after using eq.~(\ref{StoT}).

Of course the presence of other interactions and charged fields does not affect the proof. The argument above can be run backwards, to show that the FI term can be reabsorbed via a field redefinition and a K\"ahler transformation. Notice that the FI term generates from the mass term $(1/2) M^2 V_m^2$ in the K\"ahler potential. Equivalently, a FI term can be reabsorbed by a field redefinition and a K\"ahler transformation only when such term is present.

\subsection{An anomaly-free model with genuine FI term}

We formulate now an explicit model that provides an existence proof of theories with the following properties: \textit{(i)} presence of a genuine FI term; \textit{(ii)} cancellation of all gauge anomalies; 
\textit{(iii)} existence of a locally stable vacuum with all scalar field stabilized at tree level; \textit{(iv)} absence of exact global symmetries; \textit{(v)} all physical masses and energy densities parametrically small with respect to the Planck scale, even when the FI term is assumed to be quantized in Planck mass units. The chosen example has also the following features: the vacuum has positive energy; the gauged $U(1)$ R-symmetry can be chosen to be either broken or unbroken on the vacuum; there exists a limit where also supersymmetry is recovered  on the vacuum.

The model contains one vector supermultiplet, associated with the $U(1)$ R-symmetry that generates the constant FI term $\xi$, and 24 chiral supermultiplets, transforming linearly under the gauged $U(1)$: one ($\Phi_+$) with charge $q_+=+ \xi$ and 23 ($\Phi_-^{i=1..23}$) of charge $q_-=- \xi$. The corresponding fermions have then charges $Q[\psi_+] = 3 \, \xi/2$ and  $Q[\psi^i_-]= - \xi/2$. It is immediate to check that the anomaly cancellation conditions of eqs.~(\ref{cubic}) and (\ref{linear}) are identically satisfied.

We discuss first the model with canonical K\"ahler potential,
\be
K_0 = \left| z_+ \right|^2 + \sum_{i=1}^{23} \left| z_-^i \right|^2  \, ,
\ee
`minimal' superpotential with the appropriate charge\footnote{In the case of a general linear superpotential, $W_0 = \sum_{i}^{23} M^2_{i} \, z_-^i $, we can always move to the form of $W_0$ given in eq.~(\ref{wsimple}) by a suitable rotation in the space of the $z_-^i$ fields, which leaves $K_0$ and $f_0$ invariant.},
\be
\label{wsimple}
W_0 = M^2 \, z_-^1 \, ,  
\ee
where $M$ is a real mass parameter, and constant gauge kinetic function,
\be
f_0 = \frac{1}{g^2} \, .
\ee
In such a case, the D-term of eq.~(\ref{Dgen}) reads
\be
D = \xi \, \left( 1 + \sum_{i=1}^{23}  \left| z_-^i \right|^2  -  \left| z_+ \right|^2 \right) \, .
\ee
The scalar potential of eq.~(\ref{vgen}) has
\be
V_F = e^{K_0} \, M^4 \, \left[ 1 +  \left| z_-^1 \right|^2 \left( \sum_{i=1}^{23}  \left| z_-^i \right|^2  +  \left| z_+ \right|^2 -1 \right) \right] \, , 
\ee
and
\be
V_D = \frac{g^2 \, \xi^2}{2}  \, \left( 1 + \sum_{i=1}^{23}  \left| z_-^i \right|^2  -  \left| z_+ \right|^2 \right)^2 \, .
\ee 
Notice that $V_F\geq M^4$ and $D\geq0$. For $M=0$, $V_F$ is identically vanishing and $V_D$ can relax to a supersymmetric Minkowski vacuum with unbroken supersymmetry and spontaneously broken $U(1)$ gauge symmetry. For $M \ne 0$, the full potential $V$ is strictly positive definite and always admits classically stable dS vacua. 

For $g \, \xi < M^2$, the $U(1)$ gauge symmetry is unbroken, $\langle W_0 \rangle = \langle z_+ \rangle =\langle z_-^i \rangle =0$, the vacuum energy density is $\langle V \rangle = M^4 + g^2 \, \xi^2 /2$ and the squared masses for the scalar fields $ (z_-^1,z_-^{2...23},z_+)$ are $(g^2 \xi^2, \ M^4 + g^2 \xi^2, \ M^4 - g^2 \xi^2)$, respectively. They are all positive and of the order of the Hubble scale.

For $g \, \xi > M^2$, the $U(1)$ gauge symmetry is spontaneously broken, the field $z_+$ develops a VEV $v$ satisfying the equation $M^4 \, e^{v^2} = g^2 \, \xi^2 \, (1 - v^2)$.  In this case the vacuum energy is 
$\langle V \rangle=\frac12 g^2 \xi^2(1-v^2)(3-v^2)$, which is always positive except for $M=0$ where it vanishes. The squared masses for the scalar fields $(z_-^1,z_-^{2...23},{\rm Re}\,z_+)$ are all positive and given by $(g^2 \xi^2(1-v^2),\ 2g^2 \xi^2(1-v^2),\ 2 g^2 \xi^2 v^2(2-v^2))$, respectively, while ${\rm Im}\, z_+$ is eaten by the vector boson, which has a mass $\sqrt2 g\xi v$. The masses are again of the order of the Hubble scale, except in the supersymmetric limit ($M\to0$ and $v\to1$), where the de~Sitter curvature goes to zero while the vector superfield remains massive, eating the chiral superfield $\Phi_+$ in a supersymmetric way.

The simple model described above has a large amount of global symmetries, since the canonical K\"ahler potential $K_0$ is invariant under $U(24) \times U(1)_R$, gauge interactions break $U(24)$ to $U(23) \times U(1) \times U(1)_R$, superpotential interactions in $W_0$ break $U(23) \times U(1) \times U(1)_R$ into $U(22)\times U(1)  \times U(1)_R^\prime$. However, it is relatively simple to break all the residual global symmetries by introducing higher-dimensional operators into the Kahler potential ($K=K_0 + \Delta K$) and the superpotential ($W =W_0 + \Delta W$), with modifications such as $\Delta K= a_{i \bar \j}\, z_-^i \, \ov{z}_-^{\bar \j} \, |z_+|^2$ and $\Delta W=b_{ij} \, z_-^i \, z_-^j z_+$. Since all the scalar fields have positive squared masses at tree level, the presence of the higher-dimensional operators does not destabilize the vacuum if their coefficients are small enough, $a_{i \bar \j},b_{ij}\ll1$.

All the physical masses and energy densities are controlled by the two parameters $M^2$ and $g\xi$.
Even if we assume that the FI term $\xi$ is quantized in units of the Planck mass, by choosing small values
for $M$ and $g$ the spectrum of the theory is parametrically below the Planck scale.
Interestingly for $\xi\sim O(1)$ in Planck units the relevant mass scale $g\xi$ 
matches the cut-off scale expected from the weak gravity conjecture (WGC) of \cite{ArkaniHamed:2006dz}.
The model above avoids violating the sharp bound from the WGC since it
 always contains at least one charged particle with mass $m \lesssim g M_{Pl}$.
However the absence of an hierarchy between the relevant mass scale of the model 
and the expected cut-off may signal some deep inconsistency at the quantum gravity level
and explain the absence of explicit string theory constructions with genuine FI terms. 
Alternatively, new physics at the scale $g\xi$ may act as an innocuous spectator, 
leaving the supergravity Lagrangian with the FI term as a consistent truncation of the whole theory. 
We are not aware of sharp arguments against any of the two possibilities.

Notice finally that our model is not in conflict\footnote{We thank Z.~Komargodski for discussions on this point.} with the results of ref.~\cite{recent}, since in the rigid limit none of the matter fields is charged under the gauged $U(1)$: the interactions of the gauged $U(1)$ are a supergravity phenomenon, as in many other examples of gauged supergravity theories.

\section{$\Ne1$ truncations of $\Ne2$ models with classically stable dS vacua}

The only known models with extended supersymmetry and classically stable dS vacua are the $\Ne2$ models constructed by Fr\'e, Trigiante and Van Proeyen (FTVP) in \cite{N2FI} and some simple $\Ne2$ extensions based on the same ingredients \cite{altri}. One of their crucial ingredients is the presence of a $\Ne2$ FI term, corresponding to an arbitrary constant in the moment map. It is interesting to study the features of the $\Ne1$ models obtained from the FTVP models by consistent truncations, to understand the relation between FI terms in $\Ne2$ and $\Ne1$ supergravity. We will focus on the first and simplest of the three FTVP examples, with three vector multiplets, a $U(1)$ FI term and no hypermultiplets. The other two examples do not add qualitatively important ingredients to the truncated $\Ne1$ theories and will not be discussed here: the interested reader will find more details in \cite{tesi-fc}.

Following \cite{N2FI}, we consider $\Ne2$ gauged supergravity with three vector multiplets and no hypermultiplets. The three complex scalar fields in the vector multiplets parameterize the special K\"ahler manifold
\be
\frac{SU(1,1)}{U(1)}\times\frac{SO(2,2)}{SO(2)\times SO(2)} \, . 
\ee
For a suitable choice of field coordinates $(S,y_0,y_1)$, the K\"ahler potential reads
\be
K=- \log (S + \overline{S}) - \log \left( \frac{Y}{2} \right) \, ,
\qquad
Y=1 - 2 (|y_0|^2+|y_1|^2) + \left|  y_0^2 + y_1^2 \right|^2 \, .
\label{Kahler2}
\ee
The gauge group is $G = SO(2,1) \times U(1)$. We denote by $e_0$ the coupling constant of the non-compact non-Abelian factor $SO(2,1)$, and by $e_1$ the parameter controlling the $\Ne2$ FI term of the compact $U(1)$ factor: it will not be restrictive to take both of them positive. 

Denoting with $A_\mu^a$ the four vector bosons [$a=1,2,3$ for $SO(2,1)$, $a=4$ for $U(1)$], the components of the four Killing vectors along the three complex scalar fields are:
\be
X_1^S= X_2^S = X_3^S = X_4^S =  0 \, , 
\ee
\be
X_1^{y_0} = - \frac{i}{2} \, e_0 \, (1+y_0^2-y_1^2) \, , 
\quad
X_2^{y_0}=\frac{1}{2} \, e_0 \, (1-y_0^2+y_1^2) \, ,
\quad
X_3^{y_0} = i \, e_0 \, y_0 \, , 
\quad
X_4^{y_0} = 0 \, ,
\ee
\be
X_1^{y_1} = - i \, e_0 \,  y_0  \, y_1 \, ,
\quad
X_2^{y_1} = - e_0 \, y_0 \, y_1 \, , 
\quad
X_3^{y_1} = i \, e_0 \, y_1 \, ,
\quad
X_4^{y_1} = 0 \, .
\ee
In $\mathcal N=2$ supergravity, the object that plays the r\^ole of FI term is the triholomorphic momentum map $\mathcal P_a^x$ ($x=1,2,3$), which is constant in the absence of hypermultiplets. In this model, it is zero for $a=1,2,3$ directions, while $\mathcal P_4^x$ is a constant tri-vector with modulus $e_1$.

In the absence of hypermultiplets, the scalar potential can be written as the sum ${\cal V} = {\cal V}_1 + {\cal V}_3$, where ${\cal V}_1$ and ${\cal V}_3$ are related with the square of the supersymmetry transformation of the gauginos and of the gravitinos, respectively. Explicitly, the two contributions to the scalar potential read:
\be
 {\cal V}_1 =  \frac{e_0^2}{2 ReS} \, \frac{P_2^+(y)}{P_2^-(y)} \, ,
\qquad
 {\cal V}_3 = \frac{e_1^2}{2 ReS} \, |\cos\theta+i\,S\sin\theta|^2 \, , 
\label{pot2}
\ee
where
\be
\label{pot2bis}
P_2^-(y) = Y \, ,
\qquad
P_2^+ (y)=1-2 |y_0|^2 + 2 |y_1|^2 + \left| y_0^2 + y_1^2 \right|^2 \, ,
\ee
and the angle $\theta$ is a constant free parameter that describes the magnetic rotation of one gauge group factor with respect to the other. In the following, it will not be restrictive to assume $0 < \theta <\pi/2$. The potential is minimized for
\be
\langle S \rangle = \frac{e_0}{e_1} \frac{1}{\sin\theta} + i \, \cot \theta \, ,
\quad 
\langle y_0 \rangle \ {\rm undetermined} \, ,  
\quad 
\langle y_1 \rangle = 0 \, ,
\ee
and the vacuum energy density is independent of $\langle y_0 \rangle$ and given by
\be
V_0 \equiv \langle {\cal V} \rangle = e_0 \, e_1 \, \sin\theta  
= e_1^2 \, \sin^2 \theta \, \langle \Re S \rangle  > 0 \, .
\ee
On the vacuum, two of the four vector fields, associated with the two non-compact generators of $SO(2,1)$, become massive, absorbing two Goldstone degrees of freedom from the scalar field $y_0$, and the other two vectors remain massless. 

The spectrum is most easily computed around the vacuum with $\langle y_0 \rangle = 0$. In such a case, the massive vectors are precisely $(A_\mu^1,A_\mu^2)$, with mass $m_V^2 = V_0/4$, whilst the massless vectors are  $(A_\mu^3,A_\mu^4)$. The two physical complex scalars $S$ and $y_1$ have masses $m_S^2 = 2 \, V_0$ and $m_1^2 = V_0$, and we are in the presence of a classically stable dS vacuum, with completely broken $\Ne2$ supersymmetry and vanishing Lagrangian mass terms for the gravitinos.

The rules for consistently truncating gauged $\Ne2$ supergravities to $\Ne1$ can be found in \cite{trunc}. We must set to zero one of the two supersymmetry transformation parameters, and project out from the $\Ne2$ gravitational multiplet one of the two gravitini and the graviphoton, to obtain the $\Ne1$ gravitational multiplet. Each of the three vector multiplets of $\Ne2$ contains one vector boson, two spin-1/2 fermions and one complex scalar, and can be truncated to either a $\Ne1$ chiral multiplet or to a $\Ne1$ vector multiplet. Starting from three vector multiplets in $\Ne2$, and applying the rules of
\cite{trunc}, we find that there are two different consistent truncations to $\Ne1$: the first with $n_V =1$ vector multiplets and $n_C = 2$ chiral multiplets; the second with $n_V =2$ vector multiplets and $n_C = 1$ chiral multiplets. Truncations with $n_V=0, n_C=3$ and $n_V=3, n_C=0$ are inconsistent, because the massive vector bosons (including the graviphoton) associated with the two non-compact generators of $SO(2,1)$ must be always truncated away, and their Goldstone degrees of freedom are contained in the complex scalar $y_0$. We now discuss the two consistent truncations in turn: because of the spontaneously broken non-compact gauge invariance, it will not be restrictive to concentrate for simplicity on the vacuum with $\langle y_0 \rangle =0$. We will see that the FI term of the truncated $\Ne1$ theory is associated,
in the $\Ne2$ theory, to the linear combination of the generator of the compact subgroup of $SO(2,1)$, if gauged, and the component of the $\Ne2$ FI terms that survive the truncation, if any. On the other hand, the other components of the $\Ne2$ FI terms, if present, produce superpotential terms in the truncated theory.

\subsection{Truncation with $n_V=1$ and $n_C=2$}

The first possibility for a consistent $\Ne1$ truncation preserves one $\Ne1$ vector multiplet, containing the vector $A_\mu^3$ associated with the compact $SO(2) \sim U(1)$ generator inside $SO(2,1)$, and two $\Ne1$ chiral multiplets, containing the scalar fields $S$ and $y_1$. The K\"ahler potential is obviously the one of eq.~(\ref{Kahler2}) evaluated for $y_0=0$. The $\Ne1$ gauge kinetic function is $f=S$, as can be read directly from the $\Ne2$ theory. For consistency, the scalar potential of the $\Ne1$ theory must also coincide with the scalar potential of eqs.~(\ref{pot2}) and (\ref{pot2bis}), evaluated for $y_0=0$. An interesting feature of the truncated $\Ne1$ theory is how such a potential is generated as the sum of an F-term contribution and a D-term contribution,
\be
V = V_F + V_D \, , 
\qquad
V_F =  {\cal V}_3 \, ,
\qquad
V_D =  \left. {\cal V}_1 \right|_{y_0=0} \, ,
\ee
generated by the superpotential (defined as usual up to an irrelevant constant phase factor)
\be
W=i \, e_1 \, y_1 \, (\cos\theta+i\,S\,\sin\theta) \, .
\ee
It is curious that the $\Ne2$ FI term associated with the $U(1)$ factor of the $\Ne2$ gauge group and with the constant $e_1$ is mapped into the $\Ne1$ F-term potential, whilst the $\Ne2$ potential term associated with the non-compact $SO(2,1)$ factor and with the non-Abelian gauge coupling constant $e_0$ generates a $\Ne1$ FI term $\xi = - e_0$ in the auxiliary field:
\be
D_3 =   i \, K_{y_1} \, X_3^{y_1} + i \, \frac{W_{y_1}}{W} \, X_3^{y_1} = - e_0 \, \frac{1 + |y_1|^2}{1 - |y_1|^2} \, . 
\ee
As expected, the $\Ne2$ vacuum and (truncated) spectrum are reproduced also in the standard $\Ne1$ formalism: supersymmetry is broken on a dS background but the $U(1)$ gauge boson has vanishing mass.

\subsection{Truncation with $n_V=2$ and $n_C=1$}

The second and last possibility for a consistent $\Ne1$ truncation preserves two $\Ne1$ vector multiplets, containing $A_\mu^3$ and $A_\mu^4$, and only one $\Ne1$ chiral multiplet, the one containing $S$. This time the $\Ne1$ K\"ahler potential is just $K = - \log \, (S + \overline{S})$, and the $\Ne1$ superpotential vanishes, $W=0$. Instead, the gauge kinetic function, which again can be read directly from the $\Ne2$ theory, takes the non-trivial form:
\be
f_{ab} = \left( \begin{array}{cc} 
S & 0 \\ 
0 & \displaystyle \frac{S}{\cos\theta \, ( \cos\theta + i \, S \, \sin\theta)}
\end{array} \right) \, ,
\qquad
(a=3,4) \, .
\ee
Again, the scalar potential of the $\Ne1$ theory must coincide with the scalar potential of eqs.~(\ref{pot2}) and (\ref{pot2bis}), evaluated for $y_0=0$ and $y_1=0$. This time, the $\Ne1$ potential is generated entirely as a D-term contribution:
\be
V = V_D =   {\cal V}_3 + \left. {\cal V}_1 \right|_{y_0=0} 
=\frac{1}{2ReS} \, \left( e_0^2 + e_1^2 \left| \cos\theta + i \, S \, \sin\theta \right|^2 \right) \, , 
\ee
thanks to a $\Ne1$  FI~term associated with each of the two $U(1)$ factors:
\be
D_3 = - e_0 \, ,
\qquad
D_4 = - e_1 \, .
\ee
In other words, both the $\Ne2$ FI term, associated with $U(1)$ and the constant $e_1$, and the other $\Ne2$ potential term, associated with the non-compact $SO(2,1)$ and the constant $e_0$, generate constant $\Ne1$ FI terms, $\xi_3 = - e_0$ and $\xi_4 = -e_1$. As required by the consistency of the truncation, the $\Ne2$ vacuum and (truncated) spectrum are reproduced also in the standard $\Ne1$ formalism, in particular supersymmetry is broken on a dS background but the two $U(1)$ gauge bosons have vanishing masses.

\subsection{Anomalies in the truncated theory}

While the original $\Ne2$ theory does not contain chiral fermions, thus all anomaly-cancellation conditions are automatically satisfied, truncating it to $\Ne1$ may give rise to an anomalous fermion spectrum. Indeed, we can easily check that this is the case for both truncations considered in the previous subsections.

In the first case, we have a single $U(1)$ and two chiral multiplets $(S,y_1)$ with charges $q^S=0$ and $q^{y_1}=e_0$, thus $Q[\psi^S]=-e_0/2$, $Q[\psi^1]=e_0/2$ and
\be
Tr\,Q^3  = \frac{1}{2} \, e_0^3 \ne 0 \, , 
\qquad
Tr \, Q = - 10 \, e_0 \ne 0 \, .
\label{antrun}
\ee 
In the second case, the only chiral multiplet $S$ is neutral under both $U(1)$ factors, thus the anomalies are the same for both and are again proportional to those in eq.~(\ref{antrun}).

Inspired by orbifold string constructions, where potential anomalies of the truncated theories are cancelled
by twisted sectors localized at the orbifold fixed points, we may think of supplementing the field content of the truncated theories by additional charged multiplets, to achieve anomaly cancellation while keeping the same vacuum. A simple addition that recovers anomaly freedom while keeping the same vacuum consists of $n_2=5$ chiral multiplets $\Phi_2^i$ of charge $q_2^i=0$  and $n_3=125$ chiral multiplets $\Phi_3^i$ of charge $q_3^i=(3/5) \, e_0$.
 
%
\subsection*{Acknowledgments}
We thank Kiwoon Choi, Gianguido Dall'Agata, Jean-Pierre Derendinger, Sergio Ferrara, Riccardo Rattazzi, Claudio Scrucca and Edward Witten for useful discussions. G.V. would like to thank the CERN Theory Unit and the Padua Theory Group for hospitality during different phases of this work.  This research was supported in part by the European Programme {\em Unification in the LHC Era}, contract PITN-GA-2009-237920 (UNILHC), by the Fondazione Cariparo Excellence Grant {\em String-derived supergravities with branes and fluxes and their phenomenological implications}, by the ERC Advanced Grant no.267985 {\em Electroweak Symmetry Breaking, Flavour and Dark Matter: One Solution for Three Mysteries (DaMeSyFla)}, by the Padova University Project CPDA105015/10. G.V. was partially supported by the ERC Advanced Grant no.228169 {\em Physics beyond the standard model at the LHC and with atom interferometers (BSMOXFORD)}.
%

%

\begin{thebibliography}{99}
%
\bibitem{FIglobal}
P.~Fayet, J.~Iliopoulos,
{\em Spontaneously Broken Supergauge Symmetries and Goldstone Spinors},
Phys.\ Lett.\  {\bf B51 } (1974)  461-464.
%
\bibitem{FIlocal}
D.~Z.~Freedman,
{\em Supergravity with Axial Gauge Invariance},
Phys.\ Rev.\  {\bf D15 } (1977)  1173;
\\
K.~S.~Stelle and P.~C.~West,
{\em Relation between vector and scalar multiplets and gauge invariance in supergravity},
Nucl.\ Phys.\  B {\bf 145} (1978) 175;
\\
R.~Barbieri, S.~Ferrara, D.~V.~Nanopoulos and K.~S.~Stelle,
{\em Supergravity, R invariance and spontaneous supersymmetry breaking},
Phys.\ Lett.\  B {\bf 113} (1982) 219;
\\
P.~Binetruy, G.~Dvali, R.~Kallosh, A.~Van Proeyen,
{\em Fayet-Iliopoulos terms in supergravity and cosmology},
Class.\ Quant.\ Grav.\  {\bf 21 } (2004)  3137-3170 [hep-th/0402046].
%
\bibitem{recent}
Z.~Komargodski, N.~Seiberg,
{\em Comments on the Fayet-Iliopoulos Term in Field Theory and Supergravity},
JHEP {\bf 0906 } (2009)  007  [arXiv:0904.1159 [hep-th]];
\\
Z.~Komargodski, N.~Seiberg,
{\em Comments on Supercurrent Multiplets, Supersymmetric Field Theories and Supergravity},
JHEP {\bf 1007 } (2010)  017  [arXiv:1002.2228 [hep-th]].
%
\bibitem{dt}
K.~R.~Dienes, B.~Thomas,
{\em On the Inconsistency of Fayet-Iliopoulos Terms in Supergravity Theories},
Phys.\ Rev.\  {\bf D81} (2010)  065023 [arXiv:0911.0677 [hep-th]].
%
\bibitem{FIquant}
N.~Seiberg,
{\em Modifying the Sum Over Topological Sectors and Constraints on Supergravity},
JHEP {\bf 1007}, 070 (2010)  [arXiv:1005.0002 [hep-th]];
\\
J.~Distler and E.~Sharpe,
{\em Quantization of Fayet-Iliopoulos Parameters in Supergravity},
Phys.\ Rev.\  D {\bf 83} (2011) 085010  [arXiv:1008.0419 [hep-th]];
\\
T.~Banks, N.~Seiberg,
{\em Symmetries and Strings in Field Theory and Gravity},
Phys.\ Rev.\  D {\bf 83}, 084019 (2011) [arXiv:1011.5120 [hep-th]];
\\
S.~Hellerman, E.~Sharpe,
{\em Sums over topological sectors and quantization of Fayet-Iliopoulos parameters},  
[arXiv:1012.5999 [hep-th]].
%
\bibitem{models}
M.~T.~Grisaru, M.~Rocek, A.~Karlhede,
{\em The Superhiggs Effect In Superspace},
Phys.\ Lett.\  {\bf B120 } (1983)  110;
\\
E.~Cremmer, S.~Ferrara, L.~Girardello, C.~Kounnas, A.~Masiero,
{\em Superhiggs Effect With Local R Symmetry And Vanishing Cosmological Constant},
Phys.\ Lett.\  {\bf B137 } (1984)  62;
\\
A.~H.~Chamseddine, H.~K.~Dreiner,
{\em Anomaly free gauged R symmetry in local supersymmetry},
Nucl.\ Phys.\  {\bf B458 } (1996)  65-89  [hep-ph/9504337];
\\
 D.~J.~Castano, D.~Z.~Freedman, C.~Manuel,
 {\em Consequences of supergravity with gauged $U(1)_R$ symmetry},
 Nucl.\ Phys.\  {\bf B461 } (1996)  50-70 [hep-ph/9507397];
\\
N.~Kitazawa, N.~Maru and N.~Okada,
{\em Dynamical supersymmetry breaking with gauged $U(1)_R$ symmetry},
Phys.\ Rev.\  D {\bf 62} (2000) 077701  [arXiv:hep-ph/9911251];
\\
N.~Kitazawa, N.~Maru and N.~Okada,
{\em Models of dynamical supersymmetry breaking with gauged $U(1)_R$ symmetry},
Nucl.\ Phys.\  B {\bf 586} (2000) 261  [arXiv:hep-ph/0003240];
\\
N.~Kitazawa, N.~Maru and N.~Okada,
{\em R mediation of dynamical supersymmetry breaking},
Phys.\ Rev.\  D {\bf 63} (2001) 015005  [arXiv:hep-ph/0007253];
\\
 G.~Villadoro, F.~Zwirner,
 {\em De-Sitter vacua via consistent D-terms},
 Phys.\ Rev.\ Lett.\  {\bf 95 } (2005)  231602 [hep-th/0508167];
\\
 H.~M.~Lee,
{\em Supersymmetric codimension-two branes and $U(1)_R$ mediation in 6D gauged supergravity},
JHEP {\bf 0805} (2008) 028  [arXiv:0803.2683 [hep-th]];
\\
 K.~Y.~Choi and H.~M.~Lee,
{\em $U(1)_R$-mediated supersymmetry breaking from a six-dimensional flux compactification},
JHEP {\bf 0903} (2009) 132  [arXiv:0901.3545 [hep-ph]];
\\
K.~Y.~Choi, E.~J.~Chun and H.~M.~Lee,
{\em Dark matter, mu problem and neutrino mass with gauged R-symmetry},
Phys.\ Rev.\  D {\bf 82} (2010) 105028  [arXiv:1002.4791 [hep-ph]].
%
\bibitem{Witten:1985bz}
  E.~Witten,
  {\em New Issues In Manifolds Of SU(3) Holonomy},
  Nucl.\ Phys.\  {\bf B268 } (1986)  79.
%
\bibitem{ArkaniHamed:2006dz}
N.~Arkani-Hamed, L.~Motl, A.~Nicolis, C.~Vafa,
{\em The String landscape, black holes and gravity as the weakest force},
JHEP {\bf 0706 } (2007)  060  [hep-th/0601001].
%
\bibitem{N2FI}
P.~Fre, M.~Trigiante, A.~Van Proeyen,
{\em Stable de Sitter vacua from N=2 supergravity},
Class.\ Quant.\ Grav.\  {\bf 19 } (2002)  4167-4194  [hep-th/0205119].
%
\bibitem{N4FI}
  G.~Villadoro, F.~Zwirner,
 {\em The Minimal $N=4$ no-scale model from generalized dimensional reduction},
 JHEP {\bf 0407 } (2004)  055  [hep-th/0406185];
\\
 J.~Schon, M.~Weidner,
 {\em Gauged $N=4$ supergravities},
 JHEP {\bf 0605 } (2006)  034  [hep-th/0602024];
\\
 J.~P.~Derendinger, P.~M.~Petropoulos and N.~Prezas,
 {\em Axionic symmetry gaugings in N=4 supergravities and their higher-dimensional origin},
 Nucl.\ Phys.\  B {\bf 785} (2007) 115 [arXiv:0705.0008 [hep-th]];
\\
G.~Dall'Agata, G.~Villadoro and F.~Zwirner,
{\em Type-IIA flux compactifications and N=4 gauged supergravities},
 JHEP {\bf 0908} (2009) 018 [arXiv:0906.0370 [hep-th]];
\\
G.~Aldazabal, W.~Baron, D.~Marques and C.~Nunez,
{\em The effective action of Double Field Theory},
arXiv:1109.0290 [hep-th].
%
\bibitem{altri}
  O.~Ogetbil,
  {\em Stable de Sitter Vacua in 4 Dimensional Supergravity Originating from 5 Dimensions},
  Phys.\ Rev.\  {\bf D78 } (2008)  105001  [arXiv:0809.0544 [hep-th]];
\\
 D.~Roest, J.~Rosseel,
 {\em De Sitter in Extended Supergravity},
Phys.\ Lett.\  {\bf B685 } (2010)  201-207 [arXiv:0912.4440 [hep-th]].
%
\bibitem{achuc}
A.~Achucarro, A.~Celi, M.~Esole, J.~Van den Bergh and A.~Van Proeyen,
{\em D-term cosmic strings from N=2 supergravity},
JHEP {\bf 0601} (2006) 102 [arXiv:hep-th/0511001].
%
\bibitem{Dbooks}
J.~Wess and J.~Bagger, {\em Supersymmetry and Supergravity},
2nd Edition, Princeton University Press, 1992.
%
\bibitem{afl}
 L.~Andrianopoli, S.~Ferrara, M.~A.~Lledo,
 {\em Axion gauge symmetries and generalized Chern-Simons terms in N = 1 supersymmetric theories},
 JHEP {\bf 0404 } (2004)  005  [hep-th/0402142];
\\
  J.~De Rydt, J.~Rosseel, T.~T.~Schmidt, A.~Van Proeyen and M.~Zagermann,
 {\em Symplectic structure of N=1 supergravity with anomalies and Chern-Simons terms},
 Class.\ Quant.\ Grav.\  {\bf 24} (2007) 5201  [arXiv:0705.4216 [hep-th]].
%
\bibitem{cremmer}
 E.~Cremmer, S.~Ferrara, L.~Girardello, A.~Van Proeyen,
 {\em Yang-Mills Theories with Local Supersymmetry: Lagrangian, Transformation Laws and SuperHiggs Effect},
Nucl.\ Phys.\  {\bf B212 } (1983)  413.
%
\bibitem{dsw}
M.~Dine, N.~Seiberg, E.~Witten,
{\em Fayet-Iliopoulos Terms in String Theory},
Nucl.\ Phys.\  {\bf B289 } (1987)  589.
%
\bibitem{Kugo}
S.~Ferrara, L.~Girardello, T.~Kugo, A.~Van Proeyen,
{\em Relation Between Different Auxiliary Field Formulations Of N=1 Supergravity Coupled To Matter},
Nucl.\ Phys.\  {\bf B223 } (1983)  191.
%
\bibitem{anom32}
S.~M.~Christensen, M.~J.~Duff,
{\em Axial and Conformal Anomalies for Arbitrary Spin in Gravity and Supergravity},
Phys.\ Lett.\  {\bf B76 } (1978)  571;
\\  
 N.~K.~Nielsen, M.~T.~Grisaru, H.~Romer, P.~van Nieuwenhuizen,
 {\em Approaches To The Gravitational Spin 3/2 Axial Anomaly},
 Nucl.\ Phys.\  {\bf B140 } (1978)  477.
 %
\bibitem{gs}
  M.~B.~Green, J.~H.~Schwarz,
  {\em Anomaly Cancellation in Supersymmetric D=10 Gauge Theory and Superstring Theory},
  Phys.\ Lett.\  {\bf B149 } (1984)  117-122.
%
\bibitem{tesi-fc}
F.~Catino, Ph.D. Thesis, in preparation.
%
\bibitem{trunc}
 L.~Andrianopoli, R.~D'Auria, S.~Ferrara,
 {\em Supersymmetry reduction of N extended supergravities in four-dimensions},
 JHEP {\bf 0203 } (2002)  025  [hep-th/0110277];  
\\
L.~Andrianopoli, R.~D'Auria, S.~Ferrara,
{\em Consistent reduction of N=2~$\rightarrow$ N=1 four-dimensional supergravity coupled to matter},
Nucl.\ Phys.\  {\bf B628 } (2002)  387-403  [hep-th/0112192].
%
\end{thebibliography}
\end{document}